\documentclass{article}

\usepackage{natbib}                           
\usepackage{amssymb, amsmath, amsthm, mathtools}
\usepackage{times}
\usepackage{bm, graphicx, xcolor, color, subfigure, booktabs}
\usepackage[utf8]{inputenc}
\usepackage{multirow}
\usepackage{url}
\usepackage[colorlinks=true,
            citecolor = blue,
            linkcolor = blue]{hyperref} 
\usepackage{chngcntr}   
\usepackage{enumitem} 
\usepackage{lineno}
\usepackage{threeparttable}
\usepackage{orcidlink}
\usepackage{listings}

\definecolor{myblue}{RGB}{0, 0, 180}      
\definecolor{mygreen}{RGB}{0, 128, 0}     

\lstdefinelanguage{Rcustom}{
  morekeywords={FALSE,TRUE},
  keywordstyle=\color{myblue},
  morestring=[b]",
  stringstyle=\color{mygreen},
  sensitive=true,
  alsoletter={.},
}

\newcommand{\e}{{\rm e}}
\newcommand{\dd}{\textrm{d}}

\newcommand{\R}{\mathbb{R}}
\newcommand{\Reais}{\mathbb{R}}

\newcommand{\BCS}{\textrm{BCS}}

\newcommand{\ind}{\mathbb{I}}
\newcommand{\diag}{\textrm{diag}}

\newtheorem{definition}{Definition}[section]

\DeclareFontFamily{U}{mathx}{\hyphenchar\font45}
\DeclareFontShape{U}{mathx}{m}{n}{
<5> <6> <7> <8> <9> <10>
<10.95> <12> <14.4> <17.28> <20.74> <24.88>
mathx10
}{}
\DeclareSymbolFont{mathx}{U}{mathx}{m}{n}
\DeclareFontSubstitution{U}{mathx}{m}{n}
\DeclareMathAccent{\widecheck}{0}{mathx}{"71}
\DeclareMathAccent{\wideparen}{0}{mathx}{"75}
\title{Flexible modeling of nonnegative continuous data: Box-Cox symmetric regression and its zero-adjusted extension}

\author{
Rodrigo M. R. de Medeiros\,\orcidlink{0000-0001-8101-3038}$^1$\footnote{Corresponding author: Rodrigo M. R. de Medeiros, email rodrigo.matheus@ufrn.br.}~ and 
Francisco F. Queiroz\,\orcidlink{0000-0001-8368-0707}$^2$\\
{\small {\em $^1$Department of Statistics, Federal University of Rio Grande do Norte, Brazil}}\\
{\small {\em $^2$Department of Statistics, University of S\~ao Paulo, Brazil}}
}
\date{}

\begin{document}
\maketitle

\begin{abstract}
\noindent The Box-Cox symmetric distributions constitute a broad class of probability models for positive continuous data, offering flexibility in modeling skewness and tail behavior. Their parameterization allows a straightforward quantile-based interpretation, which is particularly useful in regression modeling. Despite their potential, only a few specific distributions within this class have been explored in regression contexts, and zero-adjusted extensions have not yet been formally addressed in the literature. This paper formalizes the class of Box-Cox symmetric regression models and introduces a new zero-adjusted extension suitable for modeling data with a non-negligible proportion of observations equal to zero. We discuss maximum likelihood estimation, assess finite-sample performance through simulations, and develop diagnostic tools including residual analysis, local influence measures, and goodness-of-fit statistics. An empirical application on basic education expenditure illustrates the models' ability to capture complex patterns in zero-inflated and highly skewed nonnegative data. To support practical use, we developed the new \texttt{BCSreg} R package, which implements all proposed methods.

    \noindent {\it Keywords: Log-symmetric distributions. Positive data. Box-Cox transformation. Zero-adjusted distributions. Basic education.} 
\end{abstract}

\section{Introduction}\label{sec:intro}

The normality assumption is common in statistical models but often fails for positive continuous data, which tend to exhibit skewness and heavy tails. A common solution is to transform the data using the Box-Cox (BC) transformation \citep{box1964}; however, it has well-known practical and conceptual limitations, such as restricted support of the transformed variable and difficulty of interpreting parameters on the original scale. A more reasonable approach is to adopt distributions defined on the positive real line that can flexibly model different shapes. Examples include the gamma, Weibull, Birnbaum-Saunders \citep{birnbaum1969}, and log-symmetric distributions \citep{vanegas2016}. Another notable example is the Box-Cox normal distribution, also known as Box-Cox Cole-Green, introduced by \citet{cole1992}. A positive continuous random variable $Y$ follows a Box-Cox normal distribution with parameters $\mu > 0$, $\sigma > 0$, and $\lambda \in \R$ if
\begin{equation}\label{eq:CG-transformation}
Z \coloneqq T(Y; \mu, \sigma, \lambda) = \left\{
\begin{array}{ll}
\dfrac{1}{\sigma \lambda} \left\{\left(\frac{Y}{\mu}\right)^\lambda - 1 \right\}, & \mbox{ if } \lambda \neq 0, \\
\dfrac{1}{\sigma} \log\left(\frac{Y}{\mu}\right), & \mbox{ if } \lambda = 0,
\end{array}
\right.
\end{equation}
has a standard normal distribution truncated at $\R\backslash A(\sigma, \lambda)$ (i.e., the support of the truncated distribution is $A(\sigma, \lambda)$), where
\begin{equation*}
A(\sigma, \lambda) = \left\{
\begin{array}{cl}
\left(-\frac{1}{\sigma \lambda}, \infty\right), & \mbox{ if } \lambda > 0, \\
\left(-\infty, -\frac{1}{\sigma \lambda}\right), & \mbox{ if } \lambda < 0, \\
(-\infty, \infty), & \mbox{ if } \lambda = 0.
\end{array}
\right.
\end{equation*}

The function $T$ reduces to the Box-Cox transformation when $\mu$ and $\sigma$ are both fixed at one. The Box-Cox normal distribution has been applied in various fields, such as to model cardiac biomarker reference values in twin pregnancies \citep{kawaguchi2025}, body mass index reference charts for patients with Turner Syndrome \citep{CARPINIDANTAS2025231}, and fetal subarachnoid space biometry  \citep{SimsekajnrA8773}. Moreover, the Box-Cox normal distribution reduces to the log-normal distribution when $\lambda$ equals zero. \citet{ferrari2017} extended this model by assuming that $Z$ follows a truncated symmetric distribution, introducing the class of Box-Cox symmetric (BCS) distributions. This extended class provides greater flexibility in modeling skewness and tail behavior while preserving interpretability in regression applications, making it a powerful tool for positive continuous data.

We use an application on household expenditures in basic education to highlight unexplored aspects of the BCS distributions in a regression context and to motivate our contributions. Expenditure patterns in basic education provide insights into socioeconomic disparities and investment behavior in human capital. The data are from the 2017–2018 Consumer Expenditure Survey (Pesquisa de Orçamentos Familiares, POF), conducted by the Instituto Brasileiro de Geografia e Estatística in Brazil\footnote{Available, in Portuguese, at \href{https://www.ibge.gov.br/estatisticas/sociais/populacao/24786-pesquisa-de-orcamentos-familiares-2.html}{IBGE POF 2017–2018}.}. The analysis focuses on households residing in the state of São Paulo, one of the most populous and economically developed regions in Latin America. This application aligns with the 4th Sustainable Development Goal of the United Nations, which aims to ensure inclusive and equitable quality education for all \citep{un_2030agenda}.

In each household, the POF designates a reference person, typically the individual responsible for financial and administrative decisions. The response variable corresponds to the total household expenditure on basic education assigned to the reference person. Expenditures include childcare, preschool, and regular primary and secondary schooling, as well as youth and adult education and supplementary (equivalency) programs at the primary and secondary levels, aggregated over the 12 months preceding the interview. Households with no reported expenditure receive a value of zero. The final dataset comprises $4{,}232$ households, and the proportion of zeros is approximately 93\%, indicating that most households reported no expenditure on basic education.

\begin{figure}[!htb]
\centering
\includegraphics[scale=0.5]{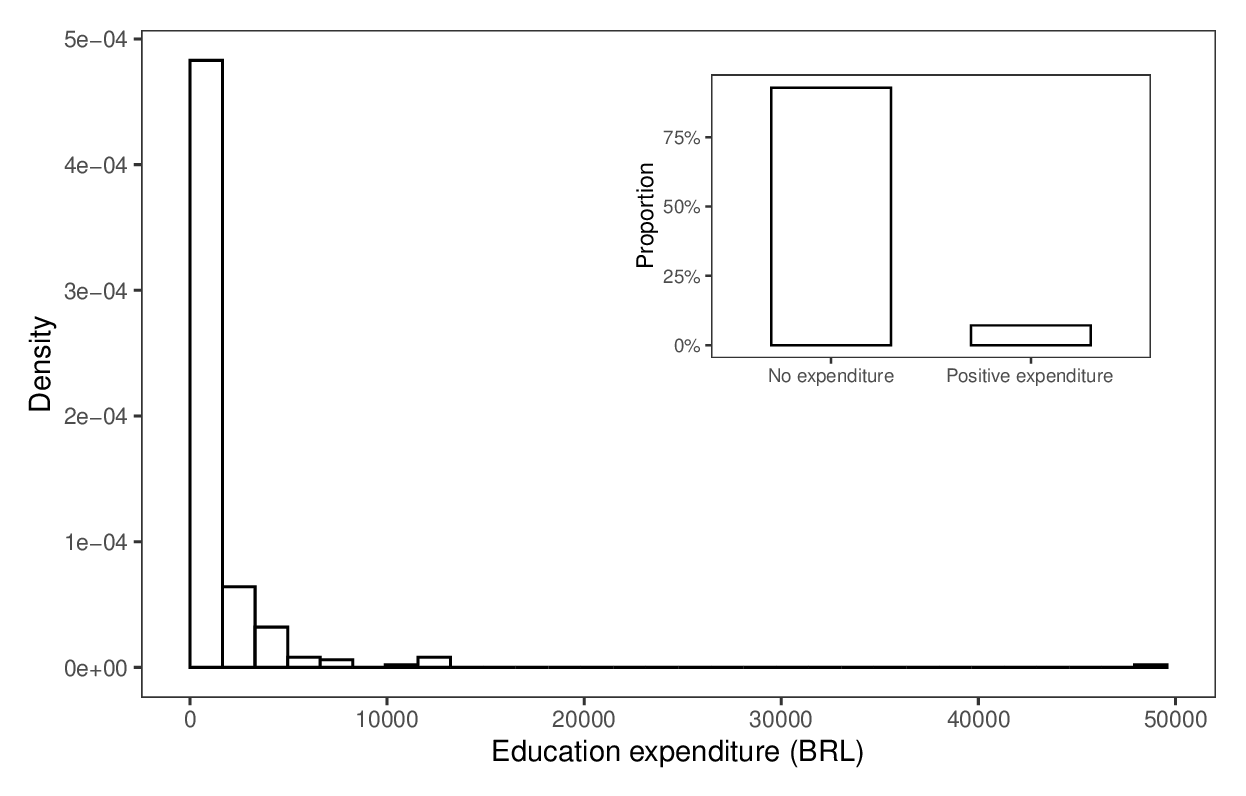}
\caption{\small Distribution of positive education expenditures. The main panel shows the density histogram for individuals with positive expenditures only, while the inset bar chart reports the proportion of individuals with zero and positive education expenditure.} \label{histogram}
\end{figure}

The distribution of education expenditures is shown in Figure~\ref{histogram}, where the histogram is based on individuals with positive expenditures only and zero expenditures are excluded. Summary statistics reveal that the median expenditure is zero, while the mean is considerably higher at BRL $108.8$, reflecting the presence of a few individuals with substantially higher spending levels. The maximum reported expenditure amounts to BRL $48{,}000$, corresponding to a 47-year-old man with two children, holding a doctoral degree and living in a household with a per capita income of BRL $6{,}520$ (total monthly disposable income of approximately BRL $40{,}600$). 

The response exhibits a highly right-skewed and heavy-tailed distribution, which motivates a regression analysis based on the BCS distributions. However, only a few specific cases of the BCS models have been explored in a regression context. The Box-Cox normal, Box-Cox power exponential, and Box-Cox $t$ regression models are popular due to their computational implementation in the R package \texttt{gamlss} \citep{rigby2005, stasinopoulos2008}. In contrast, regression models based on other BCS distributions remain unexplored in the statistical literature. Moreover, no extension of the BCS class has been proposed to explicitly account for the presence of zeros, a key feature of the data.

A natural approach for modeling positive data that include zeros is to use a mixed discrete-continuous distribution, with a probability mass at zero and an absolutely continuous distribution defined on the positive real line \citep{aitchison1955}. These probability models are often referred to as zero-adjusted or semi-continuous distributions. For instance, zero-adjusted distributions have been considered in a regression context using the gamma (see, for instance, \citet{stasinopoulos2023}), inverse Gaussian \citep{heller2006}, Birnbaum–Saunders \citep{tomazella2019}, and log-symmetric distributions \citep{cunha2024}.

This work defines a broad class of regression models for the analysis of independent positive data based on the BCS distributions and their zero-adjusted extension. The proposed framework provides a diverse set of flexible regression models suitable for a wide range of practical applications. Specifically, our contributions include: (i) the formalization of the class of BCS regression models; (ii) the introduction of zero-adjusted BCS distributions and their associated regression models; (iii)  the proposal of diagnostic methods to assess goodness-of-fit in practical applications; (iv) the development of an R package for fitting both the BCS and the zero-adjusted BCS regression models; and (v) an application on household expenditures in basic education. 

The remainder of this paper is organized as follows. Section \ref{sec:BCS_distributions} reviews the definition of the BCS distributions and introduces the class of zero-adjusted BCS distributions. Section \ref{sec:regression_models} formalizes the class of BCS regression models and extends it to accommodate a mass point at zero. It also discusses maximum likelihood estimation. Section \ref{sec:diagnostic} proposes diagnostic tools for assessing goodness-of-fit in the proposed models. Section \ref{sec:package} introduces a new R package with the computational implementations of the proposed models. 
Section \ref{sec:applications} illustrates the practical applicability of the proposed models using basic education expenditure data. Section \ref{sec:conclusions} presents final discussions, summarizes the main contributions of the paper, and outlines potential directions for future research.

\section{The Box-Cox symmetric distributions}\label{sec:BCS_distributions}

The Box-Cox symmetric distributions are defined as follows.

\begin{definition}\label{def:BCS-distributions}\textbf{(Box-Cox symmetric distributions)}.
A positive continuous random variable $Y$ is said to follow a Box-Cox symmetric (BCS) distribution with parameters $\mu > 0$, $\sigma > 0$, and $\lambda \in \R$, and density generating function (DGF) $r$, if the transformed variable $Z = T(Y; \mu, \sigma, \lambda)$, as defined in \eqref{eq:CG-transformation}, has a truncated standard symmetric distribution with probability density function (PDF) given by
$$
f_Z(z; \mu, \sigma) = \dfrac{r\left(z^2\right)}{\int_{A(\sigma, \lambda)} r\left(z^2\right) \dd z}, \quad z \in A(\sigma, \lambda),
$$
where $r:[0,\infty) \longrightarrow [0, \infty)$ satisfies $\int_0^\infty u^{-1/2}r(u)\dd u = 1$. We write $Y \sim \BCS(\mu, \sigma, \lambda; r)$.
\end{definition}

The PDF of $Y$ can be written as
\begin{equation}\label{eq:BCS_PDF}
f(y; \mu, \sigma, \lambda) = \left\{\begin{array}{ll}
\dfrac{y^{\lambda-1}}{\mu^\lambda \sigma} \dfrac{r(z^2)}{R\left(\frac{1}{\sigma |\lambda|}\right)}, & \mbox{ if } \lambda \neq 0,\\
\dfrac{1}{y\sigma} r(z^2), & \mbox{ if } \lambda = 0,
\end{array}\right. \quad y > 0,
\end{equation}
    where $R(x) = \int_{-\infty}^x r(u^2)\dd u$, $x \in \R$, and $z = T(y; \mu, \sigma, \lambda)$. The cumulative distribution function (CDF) of $Y$ is given by
\begin{equation}\label{eq:BCS_CDF}
F(y; \mu, \sigma, \lambda) = \left\{\begin{array}{ll}
\dfrac{R(z)}{R\left(-\frac{1}{\sigma \lambda}\right)}, & \mbox{ if } \lambda < 0,\\ 
R(z), & \mbox{ if } \lambda = 0,\\ 
\dfrac{R(z) - R\left(-\frac{1}{\sigma \lambda}\right)}{R\left(\frac{1}{\sigma \lambda}\right)}, & \mbox{ if } \lambda > 0,
\end{array}\right.
\end{equation} 
if $y > 0$; otherwise, $F(y; \mu, \sigma, \lambda) = 0$.

The DGF $r$ specifies the distribution of $Y$ within the class of the BCS probability models. Table \ref{tab:BCS_distributions} presents the density generating functions of some distributions in the BCS class. New BCS models can also be constructed from other truncated standard symmetric distributions. Note that some density generating functions depend on an additional parameter, which is denoted by $\zeta$. 
\begin{table}[h!]
    \centering
    \caption{Density generating function $r(u), u \geqslant 0$, and notations for some BCS distributions. The parameter space of $\zeta$ is $(0, \infty)$ for all density generators in the table, except for the BCPE, where $\zeta$ takes values in $[1, \infty)$.}
    \label{tab:BCS_distributions}
    \resizebox{\linewidth}{!}{
    \begin{tabular}{lll}
    \hline
    Distribution & $r(u)$ &  Notation\\
    \hline
    BC normal & $(2\pi)^{-1/2}\exp(-u/2)$  & BCNO\\
    BC $t$ & $\{B(1/2, \; \zeta /2)\}^{-1}\zeta ^{\zeta /2} (\zeta  + u)^{-(\zeta +1)/2}$  & BCT\\
    BC power exponential & $[\zeta  /\{p(\zeta )2^{1+1/\zeta }\Gamma(1/\zeta )\}]$ $\exp\left[-u^{\zeta /2}/\right.$ $\left.\{2p(\zeta )^\zeta \}\right]$ & BCPE\\
    BC logistic type I & $c\; \e^{-u}(1 + \e^{-u})^{-2}$ & BCLOI\\
    BC logistic type II & $\e^{-\sqrt{u}}\left(1 + e^{-\sqrt{u}}\right)^{-2}$ & BCLOII \\
    BC hyperbolic & $\{2 K_1(\zeta )\}^{-1}\exp\left(-\zeta  \sqrt{1 + u}\right)$  & BCHP\\
    BC slash & $\begin{cases}
        \Gamma(\zeta  + 0.5, 0.5u) \zeta  2^{\zeta } / \left(\sqrt{\pi} u^{\zeta  + 0.5}\right), & \textrm{if } u > 0,\\
        \zeta  / \left\{(\zeta  + 0.5)\sqrt{2 \pi}\right\}, & \textrm{if } u = 0.
    \end{cases}$ & BCSL\\
    BC sinh-normal & $r(z) = 1/(\zeta \sqrt{2 \pi}) \cosh(z^{1/2}) \exp\left[ - 2/\zeta^2 \sinh^2 (z^{1/2}) \right]$ & BCSN\\
    \hline
    \end{tabular}    
    }
    \begin{tablenotes}
        \footnotesize
       \item $B(\cdot, \cdot)$ and $\Gamma(\cdot, \cdot)$ are the beta and lower incomplete gamma functions, respectively, $K_1(\cdot)$ denotes the modified Bessel function of second kind, and $\sinh(\cdot)$ and $\cosh(\cdot)$ represent the hyperbolic sine and cosine functions, respectively. Furthermore, $p(\zeta) = 2^{-1/\zeta}$ $\Gamma(1/\zeta)^{1/2}$ $\{\Gamma(3/\zeta)\}^{-1/2}$ and $c \approx 1.484300029$ is a normalizing constant, obtained from the relation $\int_{0}^\infty u^{-1/2} r (u) \dd u = 1$ for the BCLOI distribution.
    \end{tablenotes}
\end{table}

The BCS distributions have some interesting properties. In particular, \citet{ferrari2017} showed that $\mu$ is a scale parameter, corresponding to the median of $Y$ when $\lambda=0$, or approximately the median when $\sigma \lambda$ is small. Moreover, all quantiles of the BCS distributions are proportional to $\mu$, regardless of the DGF $r$, which is an attractive feature for regression modeling. The parameters $\sigma$ and $\lambda$ are regarded as relative dispersion and skewness parameters, respectively. A detailed construction of the BCS distributions and a discussion of their parameter interpretation are provided in Sections 1 and 2 of the Supplementary Material, respectively. Furthermore, a particular case of the BCS distributions is the log-symmetric distributions, obtained when $\lambda=0$.

The zero-adjusted BCS distributions are constructed from a mixture of a degenerated random variable at zero and a BCS-distributed random variable. 

\begin{definition}\label{def:IBCS-distributions}\textbf{(Zero-adjusted Box-Cox symmetric distributions)}.
A random variable $\mathcal{Y}$ is said to follow a zero-adjusted Box-Cox symmetric (ZABCS) distribution with parameters $\alpha \in (0,1)$, $\mu > 0$, $\sigma > 0$, and $\lambda \in \R$ and DGF $r(\cdot)$, if its CDF is given by
\begin{align*}
F^{(0)}(y; \alpha, \mu, \sigma, \lambda) =  \alpha \ind(y \geqslant 0) + (1-\alpha) F(y; \mu, \sigma, \lambda),
\end{align*}
where $\ind(y \in A)$ denotes the indicator function, which equals $1$ if $y \in A$ and $0$ otherwise, and $F(y; \mu, \sigma, \lambda)$ is the CDF of a BCS distribution with parameters $\mu$, $\sigma$, and $\lambda$ and DGF $r(\cdot)$, given by \eqref{eq:BCS_CDF}. We write $\mathcal{Y} \sim \text{ZABCS}(\alpha, \mu, \sigma,$ $\lambda; r)$.
\end{definition}

Note that the zero-adjusted BCS distributions are not absolutely continuous, since they have a point mass at zero. Additionally, $\alpha= \mathbb{P}(\mathcal{Y}=0)$ represents the probability of observing zero, and $\mu$, $\sigma$, and $\lambda$ are the parameters of the continuous part (i.e., the BCS distributions), and can be regarded as the scale, relative dispersion, and skewness of the conditional distribution of $\mathcal{Y}$ given that $\mathcal{Y}>0$. The probability density function of  $\mathcal{Y}$ is given by\footnote{The probability measure $\mathbb{P}$ correspondent to $F^{(0)}(y; \cdot)$ is defined on the measurable space $([0,\infty), \mathcal{B})$, where $\mathcal{B}$ is the class of all Borelian subsets of $[0,\infty)$. This measure satisfies $\mathbb{P} \ll \varsigma + \delta_0$, where $\varsigma$ is the Lebesgue measure and $\delta_0$ is the Dirac measure at zero, defined by $\delta_0(A) = 1$, if $0 \in A$ and $\delta_0(A) = 0$, otherwise, for any $A \in \mathcal{B}$.}
\begin{equation} \label{betai}
f^{(0)}(y;\alpha, \mu, \sigma, \lambda)  = \left\{
\begin{array}{rcl}
\alpha,&  y=0,\\
(1-\alpha)f(y; \alpha, \mu, \sigma, \lambda), & y > 0,\\
\end{array}
\right.
\end{equation}
where $\alpha \in (0,1)$, $\mu > 0$, $\sigma > 0$, and $\lambda \in \R$ and $f(y; \mu, \sigma, \lambda)$ is given by \eqref{eq:BCS_PDF}. Depending on the distribution assigned to the continuous part, a different ZABCS distribution is obtained. Some examples are the zero-adjusted Box-Cox normal (ZABCNO), zero-adjusted Box-Cox $t$ (ZABCT), and the zero-adjusted Box-Cox power exponential (ZABCPE).

\section{Box-Cox symmetric regression models}\label{sec:regression_models}


The Box-Cox symmetric regression models are defined as follows. Let $Y_1,\ldots,Y_n$ be $n$ independent random variables, where $Y_i \sim \text{BCS}(\mu_i, \sigma_i, \lambda; r)$ for some DGF $r$, $i = 1,\ldots, n$. Consider the following regression structure:
\begin{align}\label{eq_linkfun}
    d_1 (\mu_i) = \bm{x_i}^{\top} \bm{\beta} = \eta_{1i} \quad \textrm{and} \quad d_2 (\sigma_i) = \bm{s_i}^{\top} \bm{\tau} = \eta_{2i},
\end{align}
where $\bm{\beta}=(\beta_1, \ldots, \beta_p)^{\top} \in \Reais^p$ and $\bm{\tau}=(\tau_1, \ldots, \tau_q)^{\top} \in \Reais^q$ are the unknown regression coefficients, which are assumed to be functionally independent and $p + q + 1 < n$. The terms $\eta_{1i}$ and $\eta_{2i}$ represent the linear predictors, while $\bm{x}_i = (x_{i1}, \ldots, x_{ip})^{\top}$ and $\bm{s}_i = (s_{i1}, \ldots, s_{iq})^{\top}$ are observations on $p$ and $q$ known independent variables, respectively. The model matrices $\textbf{X}=[\bm{x}_1,\ldots,\bm{x}_n]^\top$ and $\textbf{S}=[\bm{s}_1,\ldots,\bm{s}_n]^\top$ are assumed to have column rank $p$ and $q$, respectively. Additionally, we assume that the link functions $d_j : (0, +\infty) \to \mathbb{R}$, $j = 1, 2$, are strictly monotone and at least twice differentiable. These functions relate the parameters $\mu_i$ and $\sigma_i$ to the linear predictors $\eta_{1i}$ and $\eta_{2i}$, respectively, such that $\mu_i = d_1^{-1}(\eta_{1i})$ and $\sigma_i = d_2^{-1}(\eta_{2i})$.

A natural choice for $d_1$ is the logarithmic link function, which provides a straightforward interpretation of the regression coefficients in terms of multiplicative effects on all quantiles of the response. Specifically, for $k \in \{1, \dots, p\}$, suppose that $x_{ik}$, $i = 1, \dots, n$, are observations of a quantitative explanatory variable $x_k$ not shared by $\sigma_i$. If all other explanatory variables related to $\mu_i$ remain unchanged, an increase of $\delta > 0$ units in $x_k$ results in a multiplicative effect on the quantiles of the response variable by a factor given by $\exp(\delta \beta_k)$. When $x_k$ is also shared by $\sigma_i$, the interpretation of $\exp(\delta \beta_k)$ is only approximately valid if $\sigma_i \lambda \approx 0$, for all $i = 1, \dots, n$. If $\lambda=0$, $\exp(\delta \beta_k)$ represents the multiplicative effect on the median response. 

The choice of the DGF $r$ specifies different BCS regression models, thus forming a broad class of regression models for the analysis of strictly positive data. This class includes some models introduced in the literature in previous studies. For instance, the BCT \citep{rigby2004} and BCPE \citep{rigby2006} regression models. It also includes the class of log-symmetric regression models \citep{vanegas2014}, which are obtained by setting $\lambda = 0$.

Let $\bm{\theta} = (\bm{\beta}^\top, \bm{\tau}^\top, \lambda)^\top \in \R^{p + q + 1}$ be the parameter vector associated with the BCS regression model. The log-likelihood function for $\bm{\theta}$ is given by $\ell(\bm{\theta}) = \sum_{i=1}^n \ell_i(\mu_i, \sigma_i, \lambda),$ where
\begin{equation}\label{eq_LogLik_ind}
\ell_i(\mu_i, \sigma_i, \lambda) = \begin{cases}
(\lambda - 1) \log y_i - \lambda \log \mu_i - \log  \sigma_i + \log r(z_i^2) - \log R\left(\dfrac{1}{\sigma_i |\lambda|}\right), & \mbox{ if } \lambda \neq 0,\\
-\log y_i - \log \sigma_i + \log r(z_i^2), & \mbox{ if } \lambda = 0,
\end{cases}
\end{equation}
with $z_i = T(y_i; \mu_i, \sigma_i, \lambda)$. The score vector of $\bm{\theta}$ is given by $\textbf{U}(\bm{\theta})=(\textbf{U}_{\bm{\beta}}^\top, \textbf{U}_{\bm{\tau}}^\top,\textbf{U}_{\lambda} )^\top$, with
\begin{align}\label{eq-scorebcs}
\textbf{U}_{\bm{\beta}} = \textbf{X}^\top \textbf{T}_1 \bm{\mu}^\star, \quad \textbf{U}_{\bm{\tau}} = \textbf{S}^\top \textbf{T}_2 \bm{\sigma}^\star , \quad \textbf{U}_{\lambda} = \textbf{1}_n^\top \bm{\lambda}^\star,
\end{align}
where $\textbf{T}_1= \diag \{ 1/ \dot{d}_1(\mu_1), \ldots, 1/ \dot{d}_1(\mu_n)  \}$, $\textbf{T}_2= \diag \{ 1/ \dot{d}_2(\sigma_1), \ldots, 1/ \dot{d}_2(\sigma_n) \}$, $\bm{\mu}^* = (\mu_1^*, \ldots, \mu_n^*)^{\top}$, $\bm{\sigma}^* = (\sigma_1^*, \ldots, \sigma_n^*)^{\top}$, $\bm{\lambda}^* = (\lambda_1^*, \ldots, \lambda_n^*)^{\top}$, $\bm{1}_n$ denotes a $n$-dimensional vector of ones, $\dot{d}_j(t) =\dd d_j(t)/\dd t$, for $j=1,2$, $\mu_i^* = -\lambda/\mu_i + [1/(\mu_i\sigma_i)](z_i\sigma_i \lambda +1 ) z_i v(z_i)$, $\sigma_i^* = -1/\sigma_i + v(z_i) z_i^2 / \sigma_i$ if $\lambda=0$, and $\sigma_i^* = -1/\sigma_i + v(z_i) z_i^2 / \sigma_i + \xi_i / (|\lambda|  \sigma_i^2)$ if $\lambda \not=0$, $v(t)=-2 r'(t^2)/ r(t^2)$, $r'(u)=\dd r(u)/\dd u$,
\[
\small
\lambda_i^* = 
\begin{cases} 
\log\left(\dfrac{y_i}{\mu_i}\right) - z_i v(z_i) \dfrac{\text{d}z_i}{\text{d}\lambda} +  \dfrac{\text{sign}(\lambda) \xi_i}{\sigma_i \lambda^2}, & \hspace{-0.5em} \lambda \neq 0, \vspace{1em} \\ 
\log\left( \dfrac{y_i}{\mu_i} \right) - z_i v(z_i) \dfrac{\text{d}z_i}{\text{d}\lambda} , & \hspace{-0.5em} \lambda = 0,
\end{cases} \quad \hspace{-0.5em}
\]
and
\[
\dfrac{\text{d}z_i}{\text{d}\lambda} = 
\begin{cases} 
\dfrac{1}{\sigma_i \lambda^2} \left\{ \left( \dfrac{y_i}{\mu_i} \right)^\lambda \left[ \lambda \log\left( \dfrac{y_i}{\mu_i} \right) - 1 \right] + 1 \right\}
, & \hspace{-0.5em} \lambda \neq 0, \vspace{1em} \\ 
\dfrac{1}{2 \sigma_i} \left[ \log\left( \dfrac{y_i}{\mu_i} \right)^2 \right], & \hspace{-0.5em} \lambda = 0.
\end{cases}
\] 
The Hessian matrix for $\bm{\theta}$ are presented in Section 3 of the Supplementary Material.

Let $\bm{\widehat{\theta}} = (\bm{\widehat{\beta}}^\top, \bm{\widehat{\tau}}^\top, \widehat{\lambda})^\top$ denote the maximum likelihood estimator of $\bm{\theta}$, where $\bm{\widehat{\beta}} = (\widehat{\beta}_1, \ldots, \widehat{\beta}_{p})^\top$ and $\bm{\widehat{\tau}} = (\widehat{\tau}_1, \ldots, \widehat{\tau}_{q})^\top$. The estimates can be obtained by simultaneously solving the nonlinear system of equation $\textbf{U}(\bm{\theta})=\bm{0}_{p+q+1}$, where $\bm{0}_{p+q+1}$ denotes a $(p+q+1)$-dimensional vector of zeros. Since there is no closed-form solution for $\bm{\widehat{\theta}}$, numerical optimization techniques such as Newton’s method or quasi-Newton algorithms are required to compute the estimates in practice. The optimization algorithms require the specification of an initial value for the iterative process, which is denoted here by $\bm{\theta}^{(0)} = ({\bm{\beta}^{(0)}}^\top, {\bm{\tau}^{(0)}}^\top, \lambda^{(0)})^\top$. We recommend setting $\bm{\beta}^{(0)} = (\bm{X}^\top \bm{X})^{-1} \bm{X}^\top \bm{\upsilon}$, where $\bm{\upsilon} = (d_1(y_1), \ldots, d_1(y_n))^\top$. For $\bm{\tau}^{(0)}$, we suggest $\bm{\tau}^{(0)}= (\tau^{(0)}_1, 0, \ldots, 0)^\top$, with $\tau^{(0)}_1 = \text{asinh}(\text{CV}_y/1.5)/\Phi^{-1}(0.75)$, where $\text{CV}_y = 0.75(Q_3-Q_1)/Q_2$. Here, $Q_1$, $Q_2$, and $Q_3$ denote the first, second and third sample quantile of $y$, respectively, $\Phi^{-1}(\cdot)$ is the quantile function of a random variable with standard normal distribution, and $\text{asinh}(\cdot)$ is the inverse hyperbolic sine function. Finally, an initial value for $\lambda$ is $\lambda^{(0)} = 0$. 

Under suitable regularity conditions \citep{fahrmeir1985}, the maximum likelihood estimator of $\bm{\theta}$, that is, $\bm{\widehat{\theta}}$, is consistent and, for a large sample size $n$,
\[
\widehat{\bm{\theta}} \stackrel{a}{\sim} \mathcal{N}_{p+q+1} \left(\bm{\theta}, ~\textbf{J}_n(\bm{\theta})^{-1}\right),
\]
where $``\stackrel{a}{\sim}"$ denotes approximately distributed, and $\textbf{J}_n(\bm{\theta})$ is the observed information matrix. This approximation for large samples is useful for obtaining standard errors and confidence intervals for the model parameters.

As mentioned earlier, some BCS distributions may include an additional parameter, denoted by $\zeta$. Following \citet{lucas1997}, this parameter (or parameter vector) is kept fixed in the estimation process. To determine an appropriate value for $\zeta$, we suggest to choose $\widehat{\zeta}$ as the solution to
\[
\widehat{\zeta} = \underset{\zeta \in \Theta^\zeta}{\mbox{argmin}} \Upsilon_\zeta,
\]
with 
\begin{align}\label{upsilonmeasure}
\Upsilon_\zeta = n^{-1} \displaystyle \sum_{i=1}^n | \Phi^{-1}[F(y^{(i)}; \widehat{\mu_i}, \widehat{\sigma_i}, \widehat{\lambda})] - \upsilon^{(i)}|,
\end{align}
where $\Theta^\zeta$ denotes the parameter space for $\zeta$, $y^{(i)}$ is the $i$th order statistic of $y$, $\upsilon^{(i)}$ corresponds to the expected value of the $i$th order statistic in a random sample of size $n$ of the standard normal distribution. Additionally, $\Phi(\cdot)$ is the CDF of the standard normal distribution. Note that $\Phi^{-1}[F(Y; \mu, \sigma, \lambda)]$ follows a standard normal distribution. The goal is to choose $\zeta$ so that $(\Phi^{-1}[F(y^{(1)}; \widehat{\mu_i}, \widehat{\sigma_i}, \widehat{\lambda})],$ $\ldots,$ $\Phi^{-1}[F(y^{(n)}; \widehat{\mu_i}, \widehat{\sigma_i}, \widehat{\lambda})])$ approximates an ordered standard normal sample. This measure was proposed by \cite{vanegas2014} and recently used in \cite{queiroz2024}. One may also determine $\zeta$ as the value that maximizes the profile log-likelihood function for $\zeta$, given by $\ell^*(\zeta) = \ell(\widehat{\bm{\theta}}_\zeta)$, where $\widehat{\bm{\theta}}_\zeta$ represents the maximum likelihood estimate of $\bm{\theta}$ for a fixed $\zeta$. In practice, the profiled log-likelihood can be evaluated on a predefined grid of values for $\zeta$, and the value corresponding to the maximum is then selected.

We now define the zero-adjusted BCS regression models\footnote{The notation used here occasionally overlaps with that of the BCS regression models; however, whenever the meanings differ, clear definitions are provided.}. Let $\mathcal{Y}_1, \ldots, \mathcal{Y}_n$ be independent random variables, where $\mathcal{Y}_i \sim \mbox{ZABCS} ( \alpha_i, \mu_i, \sigma_i, \lambda; r)$, for $i=1, \ldots, n$. We assume that
\begin{align}\label{ligacaoPLI}
d_0 (\alpha_i) = \bm{\mathcal{Z}_i}^{\top} \bm{\kappa} = \eta_{0i}, \quad d_1 (\mu_i) = \bm{\mathcal{X}_i}^{\top} \bm{\beta} = \eta_{1i}, \quad d_2 (\sigma_i) = \bm{\mathcal{S}_i}^{\top} \bm{\tau} = \eta_{2i},
\end{align}
where $\bm{\kappa}=(\kappa_1, \ldots, \kappa_m)^{\top} \in \Reais^m$, $\bm{\beta}=(\beta_1, \ldots, \beta_p)^{\top} \in \Reais^p$, $\bm{\tau}=(\tau_1, \ldots, \tau_q)^{\top} \in \Reais^q$, and $\lambda>0$ are the unknown parameters, $\bm{\mathcal{Z}}_i = (\mathcal{Z}_{i1}, \ldots, \mathcal{Z}_{im})^{\top}$, $\bm{\mathcal{X}}_i = (\mathcal{X}_{i1}, \ldots, \mathcal{X}_{ip})^{\top}$, and $\bm{\mathcal{S}}_i = (\mathcal{S}_{i1}, \ldots, \mathcal{S}_{iq})^{\top}$ are the observations of the covariates with $m+p+q+1 < n$, and $\eta_{0i}$, $\eta_{1i}$, and $\eta_{2i}$ are the linear predictors. The model matrices $\bm{\mathcal{Z}} = [\bm{\mathcal{Z}}_1, \ldots, \bm{\mathcal{Z}}_n]^{\top}$, $\bm{\mathcal{X}} = [\bm{\mathcal{X}}_1, \ldots, \bm{\mathcal{X}}_n]^{\top}$, and $\bm{\mathcal{S}} = [\bm{\mathcal{S}}_1, \ldots, \bm{\mathcal{S}}_n]^{\top}$ have column rank $m$, $p$, and $q$, respectively. 

The link functions $d_0:(0,1)\rightarrow \Reais$, and $d_1, d_2:(0,\infty)\rightarrow \Reais$ are strictly monotonic and twice differentiable. A natural choice for $d_0$ is the logit function, $d_0(x) = \log[x / (1 - x)], \; x \in (0, 1)$, which allows for a coefficient interpretation analogous to that in logistic regression, in terms of odds ratios. Specifically, for $j \in \{1, \dots, m\}$, let $\mathcal{Z}_{ij}$, $i = 1, \dots, n$, denote observations of a quantitative explanatory variable $\mathcal{Z}_j$. If all other covariates related to $\alpha_i$ are held constant, an increase of $\delta > 0$ units in $\mathcal{Z}_j$  leads to a multiplicative change of $\exp(\delta \kappa_j)$ in the odds ratio of observing zero. The logarithmic function remains a natural choice for $d_1$ since it preserves the interpretation of the regression coefficients discussed earlier, with $\mu_i$ being the scale parameter of the conditional distribution of $\mathcal{Y}_i$ given that $\mathcal{Y}_i > 0$.

The zero-ajusted BCS regression models have some particular cases. The BCS regression model is a limiting case when $\alpha_i = \alpha \rightarrow 0$, for all $i=1, \ldots, n$. Furthermore, the class of zero-ajusted log-symmetric regression models \citep{cunha2024} is obtained by setting $\lambda=0$. 

The likelihood function of $\bm{\theta}=(\bm{\kappa}^{\top}, \bm{\beta}^{\top},\bm{\tau}^{\top}, \lambda)^{\top}$ can be written as
\[
L(\bm{\theta}) = \prod_{i=1}^{n} f^{(0)}(y_i;\alpha_i, \mu_i, \sigma_i, \lambda) = L_1(\bm{\kappa}) L_2(\bm{\beta}, \bm{\tau}, \lambda),
\]
where
\[
L_1(\bm{\kappa}) = \prod_{i=1}^{n} \alpha_i^{\ind(y_i = 0)} (1-\alpha_i)^{1-\ind(y_i = 0)}, \quad  \quad
L_2(\bm{\beta}, \bm{\tau}, \lambda) =  \prod_{i \in \wp} f(y_i;\mu_i, \sigma_i, \lambda),
\]
and $\wp = \{i: y_i>0\}$. Note that $L(\bm{\theta})$ can be decomposed into two distinct terms: one that depends solely on $\bm{\kappa}$ and another that involves only $(\bm{\beta}^{\top},\bm{\tau}^{\top}, \lambda)^{\top}$. Consequently, the parameter vector is separable, allowing the maximum likelihood estimates to be obtained independently for $\bm{\kappa}$ (discrete component) and $(\bm{\beta}^{\top},\bm{\tau}^{\top}, \lambda)^{\top}$ (continuous component). Furthermore, $L_1(\bm{\kappa})$ is the likelihood function of a generalized linear model for binary response \citep{mccullagh1989generalized}, where the response variable is $\ind(\mathcal{Y}_i = 0)$, while $L_2(\bm{\beta}, \bm{\tau}, \lambda)$ is the likelihood function of a BCS regression model, where the response variable is restricted to observations greater than zero.

The log-likelihood function of $\bm{\theta}$ is
\begin{equation*}
\ell (\bm{\theta}) = \ell_1(\bm{\kappa}) + \ell_2(\bm{\beta}, \bm{\tau}, \lambda), \label{logverPLI}
\end{equation*}
where
\[
 \ell_1(\bm{\kappa}) = \sum_{i=1}^{n} \ell_i(\alpha_i) \quad \text{and} \quad
 \ell_2(\bm{\beta}, \bm{\tau}, \lambda) = \sum_{i \in \wp}\ell_i (\mu_i,\sigma_i, \lambda),
\]
in which $\ell_i(\alpha_i)  = \ind(y_i = 0) \log (\alpha_i) + [1-\ind (y_i = 0)] \log (1-\alpha_i)$, and $\ell_i(\mu_i, \sigma_i, \lambda)$ is given by \eqref{eq_LogLik_ind}. The score vector is given by $\textbf{U}(\bm{\theta})=(\textbf{U}_{\bm{\kappa}}^\top, \textbf{U}_{\bm{\beta}}^\top, \textbf{U}_{\bm{\tau}}^\top,\textbf{U}_{\lambda} )^\top$, where
\begin{align*}
\textbf{U}_\kappa = \bm{\mathcal{Z}}^{\top} \textbf{A} \textbf{T}_0 \bm{\alpha}, \quad \textbf{U}_{\bm{\beta}} = \bm{\mathcal{X}}^\top \textbf{T}_1^\dagger \bm{\mu}^{\dagger}, \quad \textbf{U}_{\bm{\tau}} = \bm{\mathcal{S}}^\top \textbf{T}_2^\dagger \bm{\sigma}^{\dagger} , \quad \textbf{U}_{\lambda} = \textbf{1}_n^\top \bm{\lambda}^{\dagger},
\end{align*}
where $\textbf{A} = \diag\{ \ind(y_1 = 0) - \alpha_1, \ldots, \ind(y_n = 0) - \alpha_n \}$, $\bm{\alpha} = (1/[\alpha_1(1-\alpha_1)], \ldots, 1/[\alpha_n(1-\alpha_n)])^\top$, $\textbf{T}_0= \diag \{ 1/ \dot{d}_1(\alpha_1), \ldots, 1/ \dot{d}_1(\alpha_n)  \}$, and $\textbf{T}_j^\dagger = \diag\{t_{j1}, \ldots, t_{jn}\}$, $j = 1, 2$, with $t_{1i} = 1/\dot{d}_1(\mu_i)$ and $t_{2i} = 1/\dot{d}_2(\sigma_i)$ if $i \in \wp$, and zero otherwise. Additionally, $\bm{\mu}^\dagger = (\mu_1^\dagger, \ldots, \mu_n^\dagger)^\top$, $\bm{\sigma}^\dagger = (\sigma_1^\dagger, \ldots, \sigma_n^\dagger)^\top$, and $\bm{\lambda}^\dagger = (\lambda_1^\dagger, \ldots, \lambda_n^\dagger)^\top$ are $n$ dimensional vectors, where $\mu_i^\dagger = \mu_i^*$, $\sigma_i^\dagger = \sigma_i^*$, and $\lambda_i^\dagger = \lambda_i^*$ if $i \in \wp$, and zero otherwise.

The observed information matrix is given in Section 3 of the Supplementary Material. Given the separability of the parameters $\bm{\kappa}$ and $(\bm{\beta},\bm{\tau}, \lambda)$, the observed information matrix $\textbf{J}_n(\bm{\theta})$ has a block-diagonal structure and the maximum likelihood estimation can be performed in two separate stages. 
\begin{enumerate}
\item Fit a generalized linear model with a binary observed response given by $\ind (y_i =  0)$, for $i=1, \ldots, n$, with link function $d_0(\cdot)$ and linear predictor $\eta_{0i}$. This step yields both point estimates and standard errors for $\bm{\kappa}$. 

\item Fit a BCS regression model using the observations $y_i$, for $i \in \wp$, along with the link functions $d_1(\cdot)$ and $d_2(\cdot)$ and the covariates vectors $\bm{\mathcal{X}}_i = (\mathcal{X}_{i1}, \ldots, \mathcal{X}_{ip})^{\top}$ and $\bm{\mathcal{S}}_i = (\mathcal{S}_{i1}, \ldots, \mathcal{S}_{iq})^{\top}$, for $i \in \wp$. This step provides the estimates and standard errors for the parameter vector $(\bm{\beta}^{\top},\bm{\tau}^{\top}, \lambda)^{\top}$, as well as the estimate for the extra parameter (if needed).  
\end{enumerate}

Under suitable regularity conditions \citep{fahrmeir1985}, the maximum likelihood estimator of $\bm{\theta}$, that is, $\bm{\widehat{\theta}}$, is consistent and $\widehat{\bm{\theta}} \stackrel{a}{\sim} \mathcal{N}_{m+p+q+1} \left(\bm{\theta}, ~\textbf{J}_n(\bm{\theta})^{-1}\right)$ for large samples. In Section 4 of the Suplementary Material, we present a Monte Carlo simulation study to assess the finite-sample performance of the maximum likelihood estimator in zero-adjusted Box–Cox regression models. We consider different models within the ZABCS class, including the ZABCNO, ZABCST, ZABCPE, and ZABCLOI. The evaluation is based on empirical bias, root mean squared error (RMSE), and empirical coverage rate of the 95\% confidence intervals. The results show that bias and RMSE decrease and the empirical coverage approaches the nominal level as the sample size increases. The estimators perform particularly well when the probability of zero is small. However, with moderate probability of zero and small sample size, the estimation of the continuous component parameters becomes less accurate, and coverage rates may fall below 95\%. These findings highlight the importance of having a sufficiently large sample size and a reasonable number of non-zero observations for reliable estimation.

\section{Model selection and diagnostic tools}\label{sec:diagnostic}

The BCS and zero-adjusted BCS distribution classes offer a wide range of probability models that can be applied in practice. Models can be selected using criteria like the Akaike Information Criterion (AIC), defined as $\textrm{AIC} = -2\ell(\hat{\boldsymbol{\theta}}) + 2r$, where $r$ is the total number of model parameters. \citet[Ch. 2]{burnham2002} suggest using the AIC differences $\Delta_m = \textrm{AIC}_m - \text{AIC}_{\min}$ over all candidate models, $m = 1, \dots, T$, where $T$ denotes the total number of models considered and $\textrm{AIC}_{\min} = \min\{\textrm{AIC}_1, \dots, \textrm{AIC}_T\}$. Models with $\Delta_m \in [0, 2]$ are strongly supported, while those with $\Delta_m > 10$ lack support and may be omitted from further consideration. Additionally, the $\Upsilon_{\zeta}$ measure defined in \eqref{upsilonmeasure} can also be considered. Models yielding smaller values of $\Upsilon_{\zeta}$ are more suitable compared to the others. 

Likelihood-based inference depends on parametric assumptions, and its accuracy can be compromised by the presence of atypical observations or model misspecification. In order to assess deviations from the assumed model and detect outliers, we consider the quantile residual \citep{dunn1996} for BCS regression models, defined as follows:
\begin{align}\label{eq-quantres}
r_i^q = \Phi^{-1}[F(Y_i; \widehat{\mu}_i, \widehat{\sigma}_i, \widehat{\lambda})], \quad i=1, \ldots, n,  
\end{align}
where $\widehat{\mu}_i = d_1^{-1}(\eta_{1i})$ and $\widehat{\sigma}_i = d_2^{-1}(\eta_{2i})$ denotes the fitted values for $\mu_i$ and $\sigma_i$, respectively. If the model is well fitted to the data, the quantile residuals are expected to have an approximately standard normal distribution. A visual inspection of the scatter plot of residuals versus $\widehat{\mu}_i$ and residuals versus indices of observations can reveal possible deviations from the model's assumptions. Normal probability plots and/or non-parametric tests can be used to evaluate the asymptotic normality of quantile residuals in practice. 

Residual analysis in zero-adjusted BCS regression models may be conducted in parts; that is, analyze the discrete and continuous components separately, possibly using distinct residuals for each. For the continuous part, the quantile residual defined in  \eqref{eq-quantres} may be used. For the discrete component, we suggest using the standardized Pearson residual, given by
\begin{align*}
r_i^p = \dfrac{\ind(Y_i = 0) - \widehat{\alpha}_i}{\sqrt{\widehat{\alpha}_i(1-\widehat{\alpha}_i)(1-\widehat{h}_{ii}^*)}}, \quad i=1, \ldots, n,
\end{align*}
where $\widehat{h}_{ii}^*$ is the $i$th diagonal element of $\textbf{H}^* = \textbf{Q}^{1/2} \bm{\mathcal{Z}}(\bm{\mathcal{Z}}^\top \textbf{Q}\bm{\mathcal{Z}})^{-1} \bm{\mathcal{Z}}^\top\textbf{Q}^{1/2}$ evaluated at the maximum likelihood estimate, where $\textbf{Q} = \diag\{ q_1, \ldots, q_n \}$, with $q_i = 1/[\alpha_i (1-\alpha_i) {\dot{d}_0}(\alpha_i)^2]$, for $i=1, \ldots, n$. If the model provides a good fit to the data, the standardized Pearson residuals are expected to have a mean close to zero and a standard deviation close to one.

The overall adequacy of the zero-adjusted BCS regression model can be assessed using the normalized randomized quantile residuals \citep{dunn1996}, defined as
\begin{align*} 
r_i^q = \left\{
\begin{array}{rcl}
\Phi ^{-1} (U_i), & & y_i =0,\\
\Phi ^{-1}[F^{(0)}(Y_i; \widehat{\alpha}_i, \widehat{\mu}_i, \widehat{\sigma_i}, \widehat{\lambda} )] ,& & y_i >0
\end{array}
\right.
\end{align*}
in which $U_i$ is a random variable with uniform distribution on $(0, \widehat{\alpha}_i]$. If the model is correctly specified, this residual has a standard normal distribution asymptotically. The randomized quantile residual may vary across realizations in practical situations, as the residual for the discrete part is obtained through randomization. Thus, we recommend at least four realizations to identify patterns in its behavior 

In regression analysis, a few observations may exert a disproportionate weight on the model fit, potentially influencing the estimate of the regression coefficients. These observations are called influential observations and may not be easily identified with a residual analysis. The analysis of local influence in statistical models, introduced by \citet{cook1986}, aims to assess the effect of small perturbations in the model formulation or the observed data. Let $\bm{\omega} \in \bm{\Omega}$ be a $k$-dimensional perturbation vector, where $\bm{\Omega} \subset \R^k$ is an open set; generally, we have $k=n$. Let $\ell(\bm{\theta}; \bm{\omega})$ denote the log-likelihood associated with the perturbed model for a given $\bm{\omega} \in \bm{\Omega}$, which is assumed to be twice continuously differentiable with respect to $(\bm{\theta}^\top, \bm{\omega}^\top)^\top$. Assume that $\bm{\Omega}$ contains a no-perturbation vector, say $\bm{\omega}_0$, such that $\ell(\bm{\theta}; \bm{\omega}_0) = \ell(\bm{\theta})$, for all $\bm{\theta} \in \R^{p+q+1}$. Furthermore, let $\widehat{\bm{\theta}}$ and $\bm{\widehat{\theta}_\omega}$ denote the maximum likelihood estimators under the postulated and perturbed models, respectively.

To evaluate the effect of minor perturbations, \citet{cook1986} proposes to analyze the local behavior of the likelihood displacement $\textrm{LD}(\bm{\omega}) = 2[\ell(\bm{\widehat{\theta}}) - \ell(\bm{\widehat{\theta}_{\omega}})]$ around $\bm{\omega}_0$, by examining the curvature of the function $\textrm{LD}(\bm{\omega}_0 + a \bm{d})$, where $a \in \R$ and $\bm{d}$ is a unit-norm direction vector. \citet{cook1986} showed that the normal curvature at the direction $\bm{d}$ is given by
$$
C_{\bm{d}} = 2|\bm{d}^\top \bm{\Delta}^\top \textbf{J}_n(\widehat{\bm{\theta}})^{-1}\bm{\Delta d}|,
$$
where $\bm{\Delta} = \partial^2 \ell(\bm{\theta}; \bm{\omega})/\partial\bm{\theta} \partial \bm{\omega}^\top$ evaluated at $\widehat{\bm{\theta}}$ and $\bm{\omega}_0$. The measure $C_{\bm{d}}$ is called the local influence of the perturbation in direction $\bm{d}$ on the estimate of $\bm{\theta}$. A measure of particular interest is the direction $\bm{d}_{\textrm{max}}$ associated with the largest curvature $C_{\bm{d}_{\textrm{max}}}$, that is equivalent to the unit-norm eigenvector associated with the largest eigenvalue of the matrix $\bm{B} = -\bm{\Delta}^\top \textbf{J}_n(\widehat{\bm{\theta}})^{-1}\bm{\Delta}$. The index plot of $|\bm{d}_\textrm{max}|$ can reveal influential observations. Another
possibility was proposed by \cite{lesaffre1998local}, named total local influence, which consists
of the construction of the index plot of $C_i = 2| \bm{\Delta}_i^\top \textbf{J}_n(\widehat{\bm{\theta}})^{-1} \bm{\Delta}_i|$, where $\bm{\Delta}_i$ is the $i$th column of $\bm{\Delta}$. This plot helps identify observations that are individually influential.

Local influence analysis can use different perturbation schemes; here, we focus on the case-weights scheme. The elements of the perturbation vector $\bm{\omega} = (\omega_1, \ldots, \omega_n)^\top$ are such that $0 \leqslant \omega_i \leqslant 1$, $i = 1, \ldots, n$, and $\bm{\omega}_0 = \bm{1}_n$. For the BCS regression models, the perturbed log-likelihood function is $\ell(\bm{\theta}; \bm{\omega}) = \sum_{i=1}^n \omega_i \ell_i(\mu_i, \sigma_i, \lambda)$ and
$\bm{\Delta} = (\bm{\Delta}_{\bm{\beta}}^\top, \bm{\Delta}_{\bm{\tau}}^\top, \bm{\Delta}_{\lambda}^\top)^\top$, where
$$
\bm{\Delta}_{\bm{\beta}}^\top = \textbf{X}^\top \widehat{\textbf{T}}_1 \widehat{\textbf{D}}_{\bm{\beta}}, \quad \bm{\Delta}_{\bm{\tau}}^\top = \textbf{S}^\top \widehat{\textbf{T}}_2 \widehat{\textbf{D}}_{\bm{\tau}}, \quad \textrm{and} \quad \bm{\Delta}_{\lambda} = (\lambda_1^\star, \ldots, \lambda_n^\star),
$$
where $\textbf{D}_{\bm{\beta}} = \diag\{\mu_1^\star, \ldots, \mu_n^\star\}$, $\textbf{D}_{\bm{\tau}} = \diag\{\sigma_1^\star, \ldots, \sigma_n^\star\}$, and quantities denoted with a hat (circumflex) represent values evaluated at the maximum likelihood estimates. For the zero-adjusted BCS regression models, the perturbed log-likelihood is given by $\ell(\bm{\theta}; \bm{\omega}) = \sum_{i=1}^n \omega_i \ell_i(\alpha_i) + \sum_{i \in \wp} \omega_i \ell_i(\mu_i, \sigma_i, \lambda)$ and $\bm{\Delta} = (\bm{\Delta}_{\bm{\kappa}}^\top, \bm{\Delta}_{\bm{\beta}}^\top, \bm{\Delta}_{\bm{\tau}}^\top, \bm{\Delta}_{\lambda}^\top)^\top$, where
$$
\bm{\Delta}_{\bm{\kappa}}^\top = \bm{\mathcal{Z}}^\top \widehat{\textbf{A}} \widehat{\textbf{T}}_0 \widehat{\textbf{D}}_{\bm{\kappa}}, \quad 
\bm{\Delta}_{\bm{\beta}}^\top = \bm{\mathcal{X}}^\top \widehat{\textbf{T}}^\dagger_1 \widehat{\textbf{D}}_{\bm{\beta}}^\dagger, \quad \bm{\Delta}_{\bm{\tau}}^\top = \bm{\mathcal{S}}^\top \widehat{\textbf{T}}^\dagger_2 \widehat{\textbf{D}}_{\bm{\tau}}^\dagger, \quad \textrm{and} \quad \bm{\Delta}_{\lambda} = (\lambda_1^\dagger, \ldots, \lambda_n^\dagger),
$$
with $\textbf{D}_{\bm{\kappa}} = \diag\{\alpha_1, \ldots, \alpha_n\}$, $\textbf{D}_{\bm{\beta}}^\dagger = \diag\{\mu_1^\dagger, \ldots, \mu_n^\dagger\}$, and $\textbf{D}_{\bm{\tau}}^\dagger = \diag\{\sigma_1^\dagger, \ldots, \sigma_n^\dagger\}$.

\section{Computational implementation: \texttt{BCSreg} package} \label{sec:package}

The \texttt{BCSreg} R package, available on the Comprehensive R Archive Network
(\href{https://CRAN.R-project.org/package=BCSreg}{CRAN}), was developed to facilitate the application of the proposed models. It supports both BCS and zero-adjusted BCS regression models and allows users to choose among several distributions for the continuous component, including the BCNO, BCT, BCPE, BCSL, BCHP, BCSN, BCLOI, and BCLOII distributions. The package includes diagnostic tools for model assessment, such as residual analysis, local influence and goodness-of-fit statistics. Estimation is performed via maximum likelihood, and users can choose either to fix the skewness parameter $\lambda$ or to estimate it from the data. In particular, setting $\lambda=0$ enables fitting log-symmetric and zero-adjusted log-symmetric regression models.

The main function for model fitting in the \texttt{BCSreg} package is \texttt{BCSreg()}, and its usage is illustrated below:
\begin{lstlisting}
BCSreg(formula, data, subset, na.action, family = "NO", zeta, link = "log", 
        sigma.link = "log", alpha.link, control = BCSreg.control(...), 
        model = FALSE, y = FALSE, x = FALSE, ...)
\end{lstlisting}
The \texttt{formula} argument, based on the \texttt{Formula} package \citep{zeileis2010}, can include up to three parts separated by the ``$|$'' operator. The first part specifies the model for the scale parameter. An optional second part defines a regression structure for the relative dispersion parameter. A third part, applicable only for zero-adjusted positive data, models the probability of zero. The specific model within the BCS or ZABCS class to be used is passed through the \texttt{family} argument. 

Once the fitting process is completed, an object of \texttt{S3} class ``\texttt{BCSreg}'' is returned. This object supports several methods. The \texttt{summary()} presents key inferential results, including coefficient estimates, asymptotic standard errors, partial Wald statistics, $p$-values, and the overall goodness-of-fit measure ($\Upsilon_\zeta$). The \texttt{plot()} method produces diagnostic and influence plots to facilitate model evaluation. The \texttt{extra.parameter()} function can be used to select the extra parameter in both BCS and ZABCS regression models, based on the $\Upsilon_\zeta$ measure and the profile log-likelihood. Finally, the \texttt{influence()} function returns the influence measures $\bm{d}_{\textrm{max}}$ and $C_i$ computed based on case-weight perturbation for both BCS and ZABCS regression models.

The \texttt{BCSreg} package was employed in both the simulation study and the applications presented in this paper. The code for the examples discussed in the following section, as well as those in Section 5 of the Supplementary Material, is available at the web-based supplementary files. 

\section{Analysis of total household expenditure on basic education}\label{sec:applications}

We now analyze the dataset introduced earlier. The response variable corresponds to the total amount spent on basic education (in BRL), and the dataset includes the following socioeconomic covariates: \texttt{age}, the individual’s age in years; \texttt{sex}, categorized as ``male'' and ``female''; \texttt{years\_sc}, the number of years of formal education completed by the reference person; \texttt{residence}, classified as ``urban'' or ``rural''; \texttt{income}, the per-capita disposable household income (in thousand BRL), calculated as the total disposable income of the consumption unit divided by the number of residents. Disposable income includes both monetary and non-monetary earnings, minus direct taxes, social contributions, and other mandatory deductions; and \texttt{children}, the number of children living in the same household, including those of the reference person and/or the spouse.

\begin{figure}[!htb]
\centering
\subfigure[]{\includegraphics[scale=0.38]{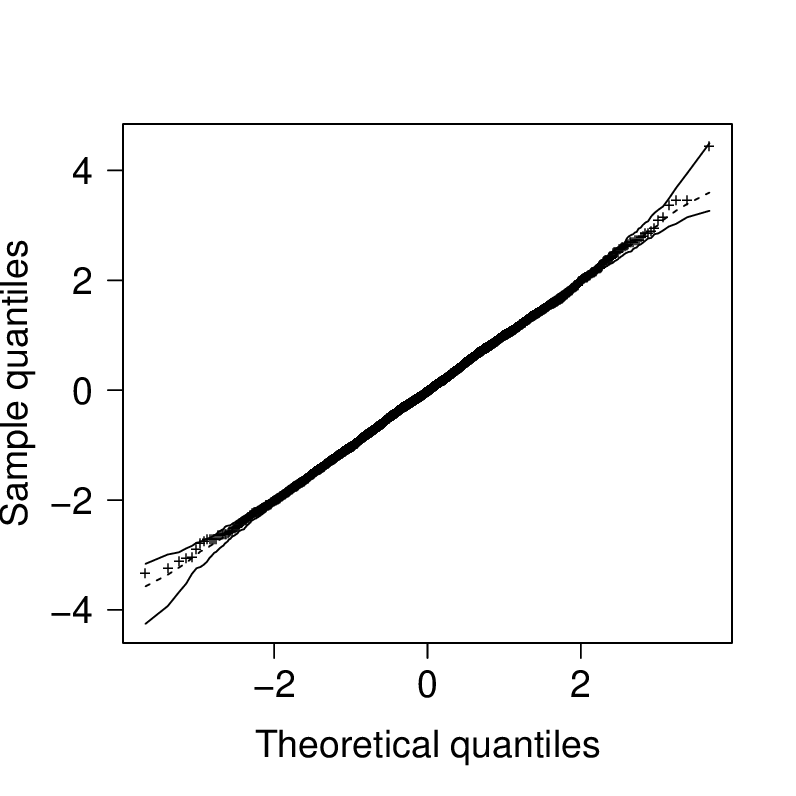}}
\subfigure[]{\includegraphics[scale=0.38]{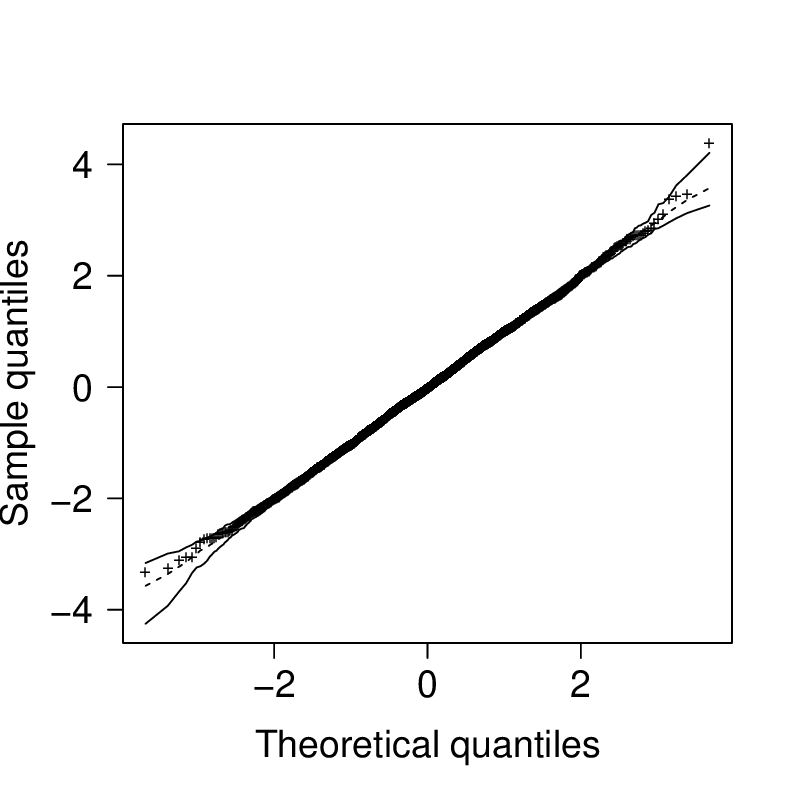}}
\subfigure[]{\includegraphics[scale=0.38]{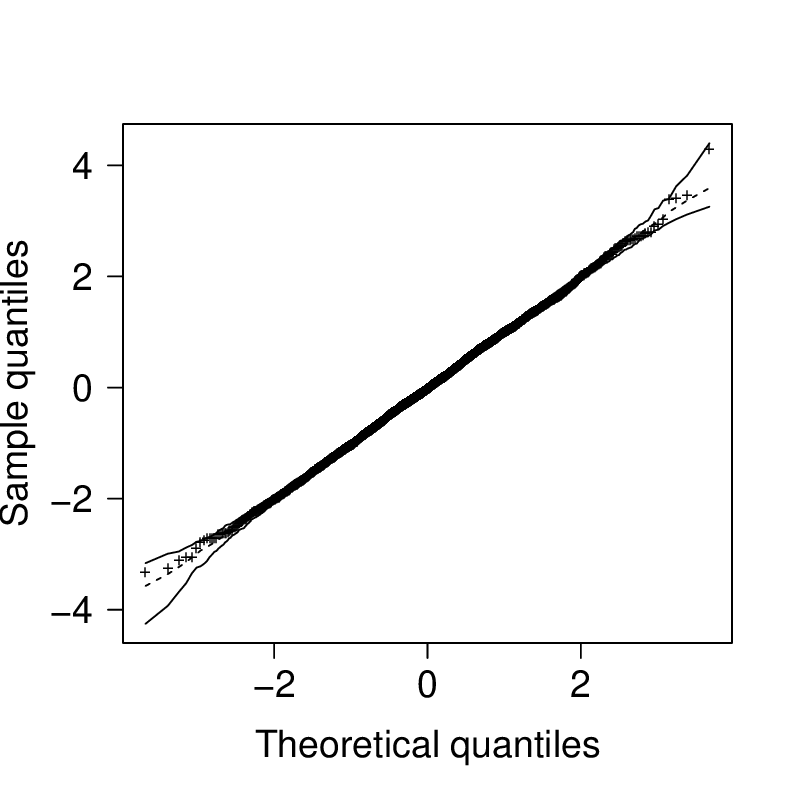}}
\subfigure[]{\includegraphics[scale=0.38]{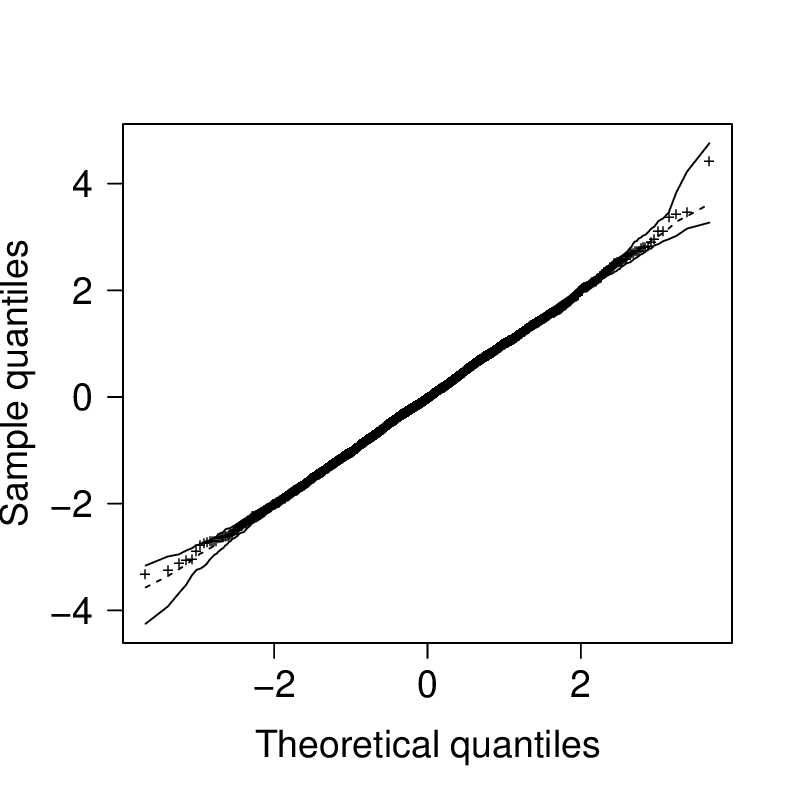}}
\caption{\small Plots of the quantile residuals with simulated envelopes for the fits under the ZABCLOII (a), ZABCHP (b), ZABCT (c), and ZABCPE (d) models.} \label{envelope}
\end{figure}

We fitted several zero-adjusted Box-Cox regression models using the R package \texttt{BCSreg}, including all covariates in the submodels for the probability of zero, conditional scale, and conditional relative dispersion. We used a logit link for $\alpha_i$ and log links for $\mu_i$ and $\sigma_i$. For comparison purposes, we also fitted regression models based on the zero-adjusted gamma (ZAGA) and zero-adjusted inverse Gaussian (ZAIG) distributions using the implementation available in the \texttt{gamlss} package \citep{stasinopoulos2008}. Table~\ref{goodness-of-fit-measures} summarizes the values of the $\Upsilon_\zeta$ measure, AIC, and AIC difference ($\Delta_m$), highlighting the ZABCLOII, ZABCHP, ZABCT, and ZABCPE models as the best-fitting candidates. The additional parameter $\zeta$ in the ZABCHP, ZABCT, and ZABCPE distributions was selected by minimizing $\Upsilon_\zeta$, yielding values of $1.2$, $4.4$, and $1.4$, respectively, which indicate heavier-tailed distributions. Figure \ref{envelope} displays the quantile residual plots with simulated envelopes for the ZABCLOII, ZABCHP, ZABCT, and ZABCPE models. Inspection of these plots indicates good agreement between the fitted models and the observed data. Among them, the ZABCLOII regression exhibited slightly better performance in the residual analysis, with all observations lying within the simulated envelope bands and requiring one fewer parameter. Therefore, it was chosen for further investigation.
\begin{table}[ht]
\centering
\caption{Goodness-of-fit measures for some models in the zero-adjusted Box-Cox symmetric class.} \label{goodness-of-fit-measures}
\resizebox{\columnwidth}{!}{
\begin{tabular}{rrrrrrrrr}
  \hline
& \multicolumn{8}{c}{Model}\\ 
\hline
Measure & ZABCLOII & ZABCHP & ZABCT & ZABCPE & ZABCNO & ZAIG & ZABCLOI & ZAGA\\ 
  \hline
$\Upsilon_\zeta$ & $0.041$    & $0.036$ &  $0.034$ &  $0.040$ &  $0.145$ &  $1.086$  & $0.207$  & $1.069$ \\
  AIC & $6573.8$   & $6574.4$  &$6574.8$  &$6576.2$  &$6611.4$ & $6611.5$ & $6653.0$ & $6717.9$\\
  $\Delta_m$ & $0.0$    & $  0.7$ &  $  1.1$ &  $  2.4$   &$37.6$   &$ 37.7$  & $ 79.2$  & $144.1$ \\
   \hline
\end{tabular}
}
\end{table}

Table~\ref{results-ZABCLOII} reports the parameter estimates, asymptotic standard errors, and the corresponding $p$-values for the ZABCLOII regression model. Before assessing the presence of potential influential observations, we note that the type of residence was not statistically significant in the zero-adjusted component; all covariates were significant in the conditional scale submodel; and only per-capita income was significant in the conditional relative dispersion submodel, at the $5\%$ significance level. The model also estimated the conditional skewness parameter $\lambda$ to be significantly different from zero, indicating that the corresponding log-logistic type-II model (obtained when $\lambda = 0$) is not reasonable for describing the data.
\begin{table}[!ht]
\centering
\caption{Estimates (Est.), standard errors (SE), and $p$-values for the ZABCLOII regression model.}\label{results-ZABCLOII}
\small
\resizebox{\columnwidth}{!}{
\begin{tabular}{rrrrrrrrrr}
\hline
& \multicolumn{3}{c}{$\log[\alpha_i/(1-\alpha_i)]$} & \multicolumn{3}{c}{$\log \mu_i$} & \multicolumn{3}{c}{$\log \sigma_i$} \\ 
\cmidrule(rr){2-4} \cmidrule(lr){5-7} \cmidrule(ll){8-10}
   & Est. & SE & $p$-value & Est. & SE & $p$-value & Est. & SE & $p$-value \\ 
  \hline
\texttt{intercept}        &  $5.416$  & $0.396$ & $<0.001$ & $5.012$  & $0.250$  & $<0.001$ & $-1.540$ & $0.360$ & $<0.001$ \\ 
\texttt{age}              &   $0.018$ & $0.005$ & $<0.001$ & $0.010$  & $0.004$  & $0.008$ & $0.006$   & $0.006$ & $0.287$ \\ 
\texttt{sex: female}      &   $0.311$ & $0.137$ & $0.024$  & $-0.212$ & $0.079$  & $0.008$ & $0.055$   & $0.115$ & $0.631$ \\ 
\texttt{years\_sc}        &  $-0.257$ & $0.022$ & $<0.001$ & $0.042$  & $0.012$  & $<0.001$ & $0.004$  & $0.015$ & $0.766$ \\ 
\texttt{residence: rural} &  $0.106$  & $0.188$ & $0.572$  & $0.205$  & $0.103$  & $0.046$ & $-0.132$  & $0.156$ & $0.396$ \\ 
\texttt{income}           &  $-6.431$ & $1.545$ & $<0.001$ & $9.658$  & $1.761$  & $<0.001$ & $4.945$  & $1.465$ & $0.001$ \\ 
\texttt{children}         &  $-0.532$ & $0.057$ & $<0.001$ & $0.265$  & $0.047$  & $<0.001$ & $0.017$  & $0.063$ & $0.788$ \\ 
$\lambda$ & $-0.291$ & $0.057$ & $<0.001$\\
  \hline
\end{tabular}
}
\end{table}

\begin{figure}[!ht]
    \centering
    \subfigure[]{\includegraphics[width=0.43\linewidth]{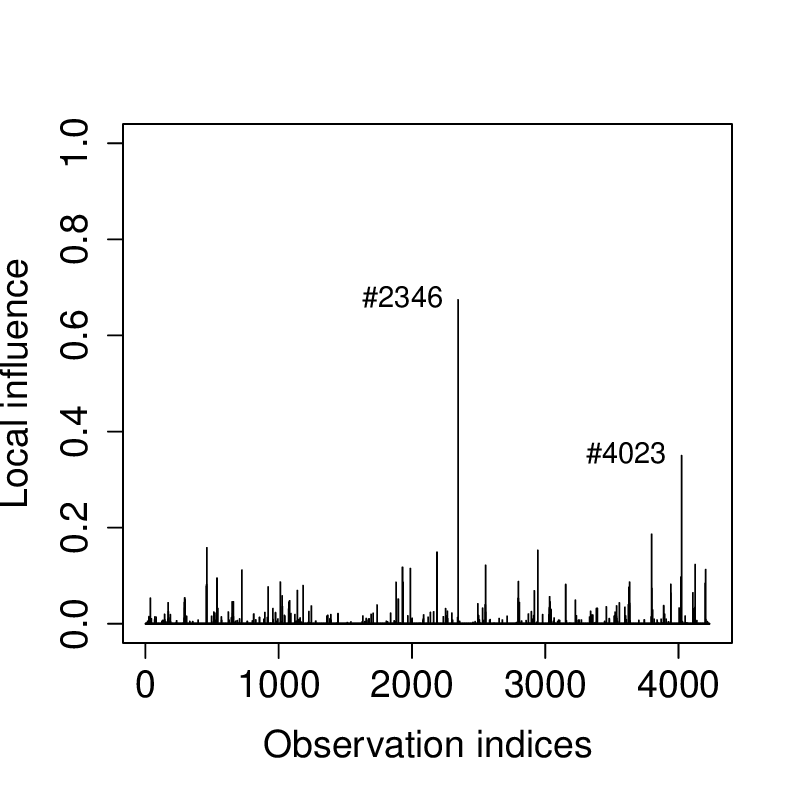}}
    \subfigure[]{\includegraphics[width=0.43\linewidth]{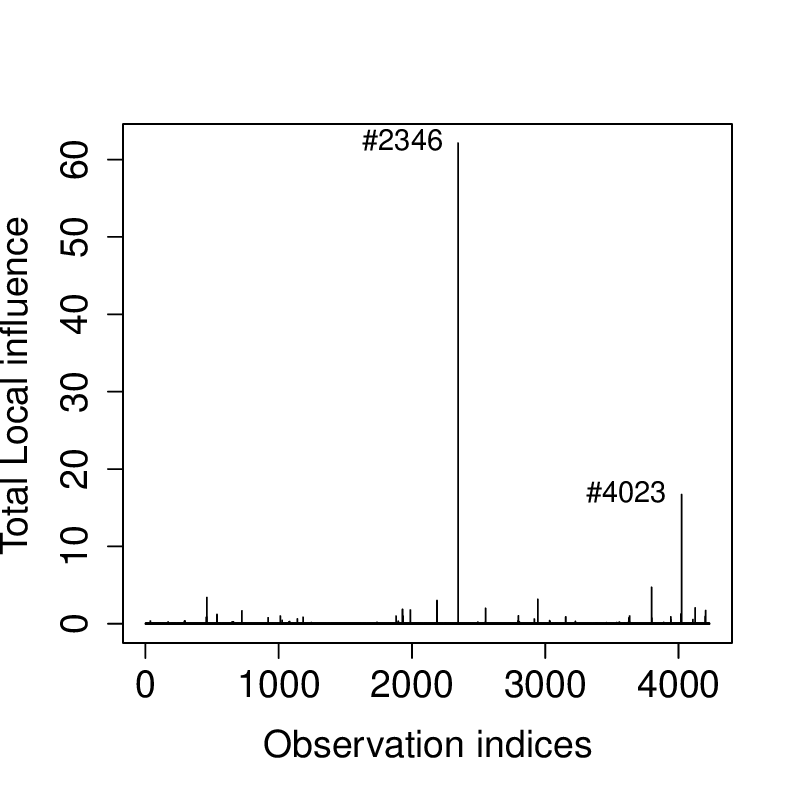}}
    \caption{Index plots of $|\bm{d}_{max}|$ (a) and $C_i$ (b).}
    \label{fig:local_influence}
\end{figure}

The index plots of $|\bm{d}_{\max}|$ and $C_i$, displayed in Figure~\ref{fig:local_influence}, indicate that observation $\#2346$ is potentially influential. This case corresponds to a 36-year-old man with $12$ years of education, residing in an urban area, with a per-capita income of BRL $39,775.76$ (the second highest) and one child. His reported expenditure on basic education was BRL $525$. An inspection of the data shows that this observation stands out among similar individuals mainly due to its unusually high income and relatively low spending on basic education. Removing observation \#2346 from the ZABCLOII model fitting affects the conditional relative dispersion submodel, most notably increasing the $p$-value for the per-capita income coefficient from $0.001$ to $0.057$. Considering the number of tests performed, this covariate can no longer be regarded as significant, confirming that observation $\#2346$ is indeed influential. The exclusion of observation $\#4023$, also highlighted in the graphs, does not cause substantial changes in the estimates and standard errors.

After variable selection, the final fitted model is presented in Table~\ref{results-ZABCLOII-final}. The likelihood ratio test comparing this model to the full model in Table~\ref{results-ZABCLOII} yields a statistic of $19.16$ on $15$ degrees of freedom ($p$-value of $0.21$), indicating that the reduced model improved the fit. For the zero-adjusted component, we find that the chance of not spending on basic education is higher for older individuals and women. Conversely, years of schooling, per-capita income, and the number of children reduce this this chance, suggesting that more educated and wealthier households, as well as those with more children, are less likely to report zero spending. For the continuous component, which models the conditional scale of basic education expenditure, age, years of schooling, per-capita income, and number of children all have positive effects, implying that older, more educated, and wealthier individuals, as well as those with more children, tend to spend more on basic education. In contrast, women spend less than men, and households in rural areas tend to spend more than those in urban areas.
\begin{table}[ht]
\centering
\caption{Estimates (Est.), standard errors (SE), and $p$-values for the final ZABCLOII regression model.}\label{results-ZABCLOII-final}
\small
\begin{tabular}{rrrrrrrrrr}
\hline
& \multicolumn{3}{c}{$\log[\alpha_i/(1-\alpha_i)]$} & \multicolumn{3}{c}{$\log \mu_i$}\\ 
\cmidrule(r){2-4} \cmidrule(r){5-7} 
   & Est. & SE & $p$-value & Est. & SE & $p$-value \\    
\hline
\texttt{intercept}        &  5.45828 & 0.38893 & $<$0.0001 &   4.96595 &  0.25857 &  $<$0.0001\\
\texttt{age}              &  0.01812 & 0.00508 & 0.00036   &   0.01030 &  0.00356 &  0.00381\\
\texttt{sex:female}       &  0.30669 & 0.13721 & 0.02541   &  $-$0.23043 &  0.08289 &  0.00543\\
\texttt{years\_sc}        & $-$0.25844 & 0.02207 & $<$0.0001 &   0.05444 &  0.01218 &  $<$0.0001\\
\texttt{residence:rural}  &          &         &           &   0.23954 &  0.11049 &  0.03016\\
\texttt{income}           & $-$0.00006 & 0.00002 & $<$0.0001 &   0.00006 &  0.00001 &  $<$0.0001\\
\texttt{children}             & $-$0.53122 & 0.05717 & $<$0.0001 &   0.27056 &  0.04888 &  $<$0.0001\\
$\sigma$         & 0.37892 & 0.01876\\
$\lambda$        & $-$0.22382 & 0.05717 & $<$0.0001\\
\hline
\end{tabular}
\end{table}

\begin{figure}[!htb]
\centering
\subfigure[\label{fig:index-res}]{\includegraphics[scale=0.36]{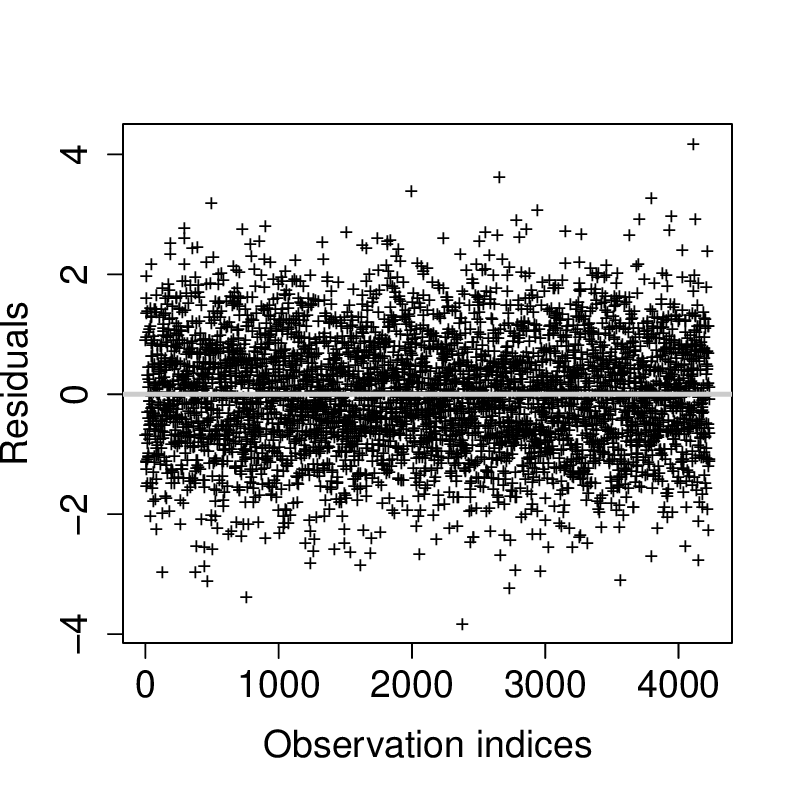}}
\subfigure[\label{fig:envelope-res}]{\includegraphics[scale=0.36]{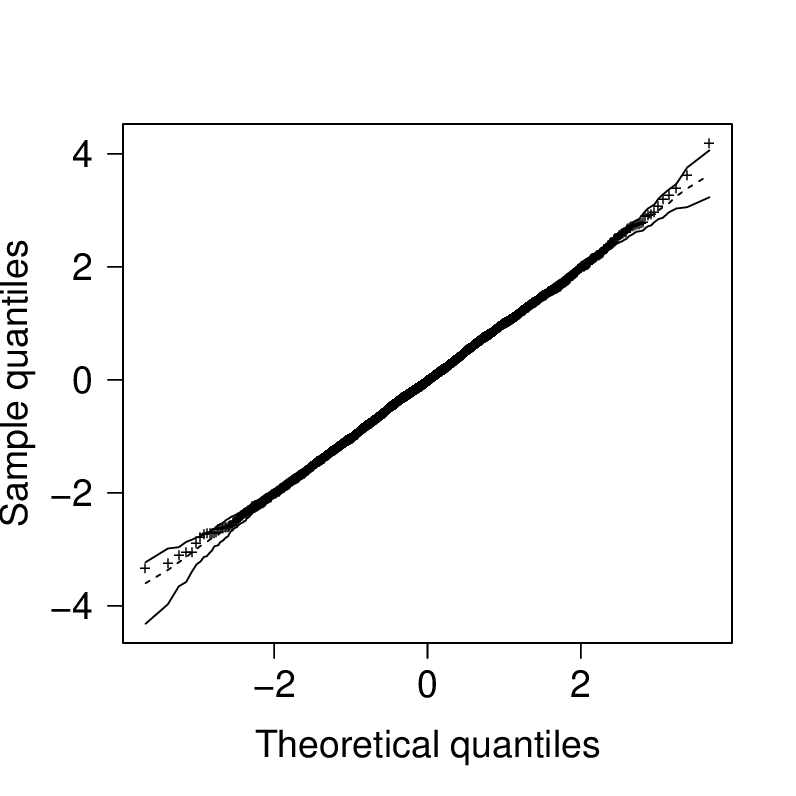}}
\subfigure[\label{fig:localinfluence}]{\includegraphics[scale=0.36]{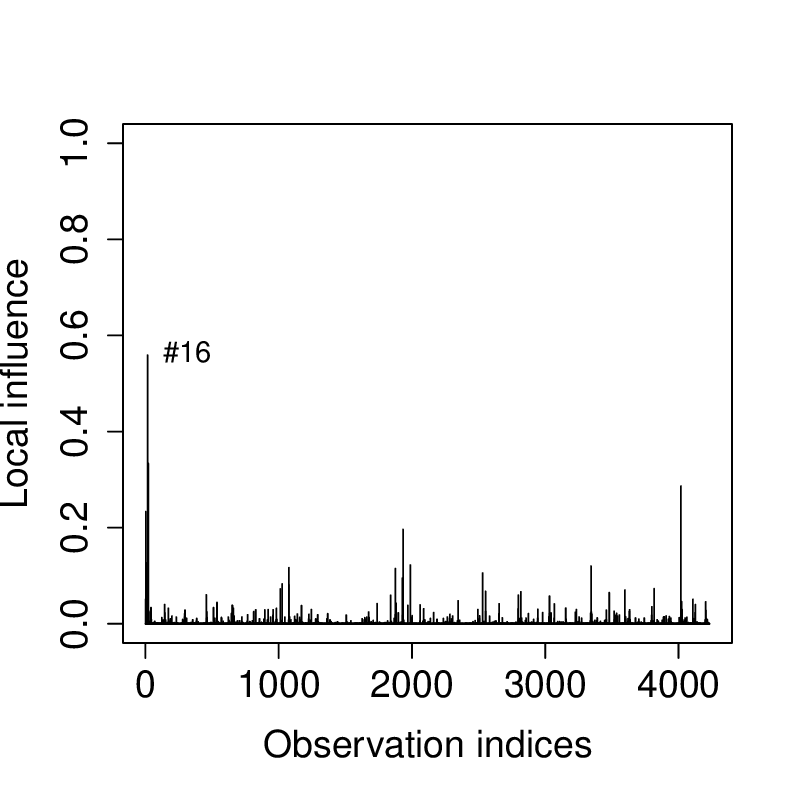}}
  \caption{Index plot of the (randomized) quantile residual (a), quantile residual with simulated envelope (b), and $|\bm{d}_\text{max}|$ (local influence) under case-weight perturbation (c).}
\end{figure}

Diagnostic plots of the final fitted model are presented in Figures \ref{fig:index-res}-\ref{fig:localinfluence}. Both the index plot of the randomized quantile residual and the quantile residual plot with simulated envelope indicate a good fit; see Figures \ref{fig:index-res} and \ref{fig:envelope-res}. The index plot of $|\bm{d}_\text{max}|$ under case-weight perturbation indicates that observation $\#16$ may be influent. To further assess its impact, the observation was removed and the model was re-estimated. However, no substantial inferential changes were observed at this time.

\section{Discussion and concluding remarks}\label{sec:conclusions}



Understanding how individuals allocate resources to education is essential to support public policy design and to monitor access to educational services. In the application presented in Section~\ref{sec:applications}, we investigated how socioeconomic characteristics influence household spending on basic education in the state of São Paulo, Brazil, using the most recent available data from the 2017 Consumer Expenditure Survey (Pesquisa de Orçamentos Familiares, POF). The results illustrate the relevance of modeling strategies capable of accommodating a large proportion of zero observations, as the data contain observational units with markedly distinct profiles. The zero-adjusted Box-Cox type II logistic regression model appears to be particularly suitable in this context, offering a better fit than competing alternatives and yielding interpretable insights into both participation and expenditure intensity.

The data on basic education expenditures contain a non-negligible proportion of zero observations, which justifies the need for a model specifically designed to handle such data characteristics. The zero-adjusted Box-Cox symmetric regression models are very flexible and allow us to model the probability of zero expenditures separately. This feature, for instance, enables the identification of variables that influence the chance of a household reporting no expenditures on basic education. To facilitate the application of these models, we provide a user-friendly implementation through the \texttt{BCSreg} R package, along with a suite of diagnostic tools and goodness-of-fit measures.

The \texttt{gamlss} package in R enables fitting the Box-Cox normal, Box-Cox $t$, and Box-Cox power exponential distributions, as well as their zero-adjusted versions through the \texttt{gamlss.inf} package. Using the \texttt{gamlss} framework, it is possible to fit these models considering structured predictors, which are naturally more flexible than those defined in \eqref{eq_linkfun} and \eqref{ligacaoPLI}. However, these packages do not allow fitting other members of the BCS class, such as the zero-adjusted Box-Cox type-II logistic model used in the analysis of basic education expenditure data. Moreover, to the best of our knowledge, neither the BCS regression models nor the zero-adjusted BCS regression models have been formally proposed in the literature or studied in terms of theoretical developments regarding inference and diagnostics. Future research will focus on semiparametric structured BCS regression models and their extensions. This will allow us to analyze basic education expenditures across different years, aiming to identify potential temporal trends as well as spatial components.

Given the flexibility of BCS and zero-adjusted BCS regression models, the broad set of tools for likelihood-based inference and diagnostic analysis, and the accessibility of user-friendly computational implementations, we support their use as reliable modeling frameworks for analyzing positive continuous data as well as mixed discrete-continuous positive data with observations at zero, respectively.

\section*{Acknowledgments}
The second author acknowledges financial support from the São Paulo Research Foundation (FAPESP), Brazil, Process Number 2024/08343-9. The authors thank Professor Silvia L. P. Ferrari for her valuable comments and suggestions on an earlier version of this manuscript.

\bibliographystyle{apalike}
\bibliography{references}

\end{document}